\documentclass{moriond}

\usepackage{cancel}
\newcommand{\doublet}[2]{ \left( \begin{array}{c}#1 \\ #2 \end{array}\right) }

\def\be{\begin{equation}}
\def\ee{\end{equation}}
\def\bea{\begin{eqnarray}}
\def\eea{\end{eqnarray}}

\newcommand{\Photo}{\includegraphics[height=35mm]{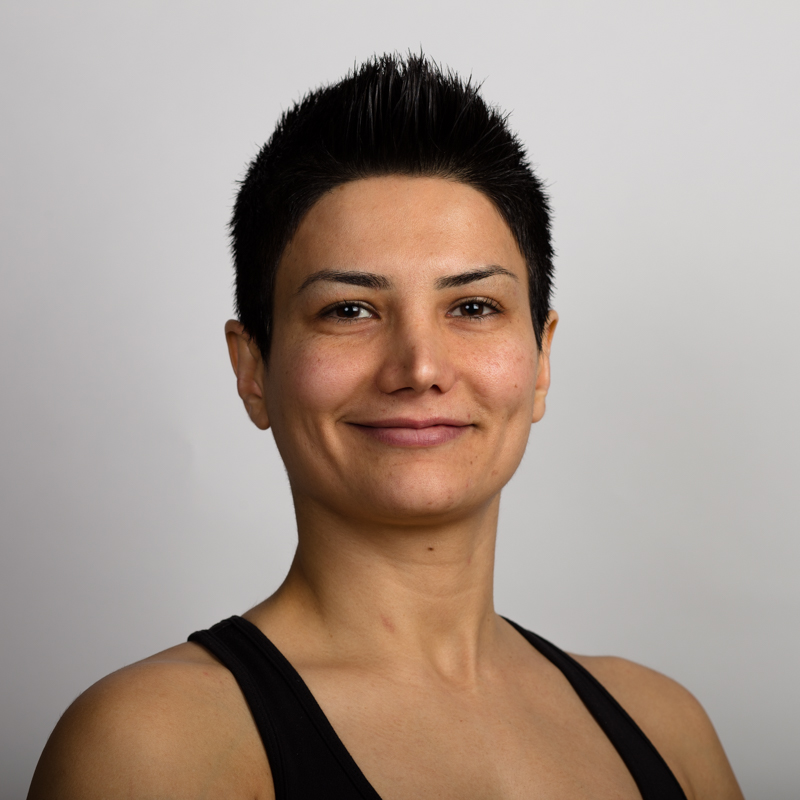}}

\begin{document}
\vspace*{4cm}
\title{CP-violation in the dark sector}

\author{Venus Keus}

\address{Department of Physics and Helsinki Institute of Physics,\\
Gustaf Hallstromin katu 2, FIN-00014 University of Helsinki, Finland}

\maketitle\abstracts{
Extended scalar sectors are a common feature of almost all beyond Standard Model (SM) scenarios which, in fact, can address many of the SM shortcomings solely on their own. While many beyond SM scenarios have lost their appeal due to the non-observation of their predicted particles or are experimentally inaccessible, scalar extensions are well within the reach of many current and upcoming experiments.
Here, we discuss the novel phenomenon of dark CP-violation which was introduced for the first time in the context of non-minimal Higgs frameworks with an extended dark sector and point out its experimental probes.}

\section{Introduction}

The Standard Model (SM) of particle physics has been extensively tested and is in great agreement with experiment with its last missing particle, the Higgs boson, discovered at the Large Hadron Collider (LHC) in 2012~\cite{Aad:2012tfa,Chatrchyan:2012ufa}. 
However, the SM falls
short of explaining several aspects of nature such as providing a viable candidate for Dark Matter (DM) to match the observational data~\cite{Ade:2015xua}. 
In the SM the fermion mass structures are given by Yukawa couplings which are parameters to be measured with no explanation from underlying physics.
Moreover,
the SM offers no explanation for the observed baryon excess in the universe.

Therefore, it is widely accepted that one needs to consider beyond SM (BSM) scenarios in pursuit of the ultimate theory of nature.
The simplest BSM scenarios aiming to conquer the SM shortcomings are non-minimal Higgs frameworks.
Extensive studies have been done to an advanced level in simple one 
singlet and one doublet scalar extensions of the SM~\cite{Englert:2011yb,Branco:2011iw}. 
These models, however, 
by construction can only partly provide a solution to the SM deficiencies.
Frameworks with a further extended scalar sector
on the other hand,
contain viable DM candidates, 
provide new sources of CP-violation and a strong first-order phase transition as essential ingredients of electroweak baryogenesis, 
contain inflaton candidates driving the inflation process in the early universe 
and provide a solution to the fermion mass hierarchy problem, all in one framework
owing to different symmetries and symmetry breaking patterns realisable in their scalar potential~\cite{Ivanov:2011ae,Keus:2013hya}.

\section{Motivations from electroweak baryogenesis and Dark Matter}

One of the most promising scenarios explaining the origin of matter-antimatter asymmetry in the universe
is electroweak baryogenesis~\cite{Morrissey:2012db} which produces the baryon excess during the electroweak phase transition. Although the SM contains all required ingredients for electroweak baryogenesis, it is unable to explain the observed baryon excess due to its insufficient amount of CP-violation~\cite{Gavela:1993ts} and the lack of a strong first-order phase transition~\cite{Kajantie:1995kf}.
Therefore, any viable electroweak baryogenesis framework requires an extended Higgs sector, predicting new scalars around the electroweak scale which directly couple to the SM Higgs boson. 
Not only such new physics is accessible at the LHC, but also gravitational waves from first-order electroweak phase transition are at the peak sensitivity of the LISA satellites~\cite{Caprini:2019egz}. 
These experimental prospects make electroweak baryogenesis very attractive compared to other baryogenesis scenarios.

On the other hand, non-minimal Higgs frameworks can naturally accommodate viable DM candidates whose stability is ensured by the symmetry of the potential or the remnants thereof after electroweak symmetry breaking. 

Simple one singlet/doublet scalar extensions of the SM, \textit{i.e.} the Higgs portal model and the 2-Higgs doublet model (2HDM), have been studied to an advanced level, even though by construction they can only partly provide a solution to these SM shortcomings and are severely constrained by experiment; They require a large Higgs-DM coupling for efficient annihilation of DM in agreement with the relic density observations, while direct and indirect DM searches demand a small Higgs-DM coupling. 
Moreover, in these models the scalar potential is inevitably CP-conserving, either by construction, 
or due to an exact discrete symmetry to stabilise the DM candidate. 
Introducing CP-violation is possible only at the expense of breaking the discrete symmetry and loosing the DM candidate as a result. What's-more, such new sources of CP-violation modify the SM-Higgs couplings and are severely constrained since they
contribute to the neutron, electron and atomic nuclei Electric Dipole Moments (EDMs)~\cite{Keus:2017ioh} whose 
measurements by the ACME collaboration are more precise than ever~\cite{Andreev:2018ayy}.

One has to go beyond simple scalar extensions to incorporate both CP-violation and DM into the model~\cite{Keus:2016orl,Keus:2019szx}. 
Specifically, if CP-violation is introduced in the extended dark sector, a novel phenomenon introduced for the first time~\cite{Cordero-Cid:2016krd} in the context of a 3-Higgs doublet model (3HDM), there will be no contributions to the EDMs and no limit on the amount of CP-violation.
It has been shown that one can construct a CP-violating DM model with unbounded dark CP-violation. 
In fact, dark CP-violating particles need not to have a Higgs-DM coupling and can interact with the SM merely through the gauge bosons, ridding the model of all current direct and indirect DM detection and LHC bounds, while yielding relic abundance 
in agreement with observation through the freeze-out or freeze-in mechanisms~\cite{Keus:2019szx}.

Moreover, owing to different symmetries that can be imposed on the scalar potential, 3HDMs with an extended dark sector allow for the exciting possibility of multi-component DM candidates and CP-violating inflation with exotic phenomenology~\cite{Hernandez-Sanchez:2020aop,Keus:2021dti}.
Such simple yet elegant frameworks can be probed by the current and upcoming DM detection experiments, for example in the nuclear recoil energy event rate measured by the XENONnT or DARWIN experiments~\cite{Aprile:2015uzo,Aalbers:2016jon}, in the photon flux excess measured by the Fermi-LAT experiment~\cite{Ackermann:2012qk}, and in the polarisations ofthe Cosmic Microwave Background accesible by the BICEP/Keck \cite{Ade:2018iql} experiments.

\section{The extended scalar potential}
We discuss the dark CP-violation phenomenon in the context of a $Z_2$-symmetric 3HDM.
A scalar potential extended by Higgs doublets, which is symmetric under a group $Z_2$ of phase rotations, can be written as the sum of two parts: $V_0$ with terms symmetric under any phase rotation, and $V_{Z_2}$ with terms symmetric under $Z_2$ \cite{Ivanov:2011ae,Keus:2013hya}, such that $V= V_0+V_{Z_2}$ where
\bea
\label{V0-3HDM}
&&V_0 = - \mu^2_{i} (\phi_i^\dagger \phi_i) + \lambda_{ii} (\phi_i^\dagger \phi_i)^2+ + \lambda_{ij}  (\phi_i^\dagger \phi_i)(\phi_j^\dagger \phi_j)
 + \lambda'_{ij} (\phi_i^\dagger \phi_j)(\phi_j^\dagger \phi_i)\,, \quad
 (i \neq j = 1,2,3), ~~
\\
\label{VZ2-3HDM}
&& V_{Z_2} =-\mu^2_{12}(\phi_1^\dagger\phi_2)+  \lambda_{1}(\phi_1^\dagger\phi_2)^2 + \lambda_2(\phi_2^\dagger\phi_3)^2 + \lambda_3(\phi_3^\dagger\phi_1)^2  + h.c., 
\eea
where the three Higgs doublets, $\phi_{1},\phi_2,\phi_3$, transform under the $Z_2$ group with the generator $g_{Z_2}=  \mathrm{\rm diag}\left(-1, -1, +1 \right)$.
The parameters of the phase invariant part, $V_0$, are real by construction. We introduce explicit CP-violation through complex parameters, $\mu^2_{12}, \lambda_1,\lambda_2, \lambda_3$ in the $V_{Z_2}$ part of the potential in Eq.~(\ref{VZ2-3HDM}).
The composition of the doublets is as follows:
\be 
\phi_1= \doublet{$\begin{scriptsize}$ H^+_1 $\end{scriptsize}$}{\frac{H_1+iA_1}{\sqrt{2}}},\quad 
\phi_2= \doublet{$\begin{scriptsize}$ H^+_2 $\end{scriptsize}$}{\frac{H_2+iA_2}{\sqrt{2}}}, \quad 
\phi_3= \doublet{$\begin{scriptsize}$ G^+ $\end{scriptsize}$}{\frac{v+h+iG^0}{\sqrt{2}}}, 
\label{explicit-fields}
\ee
where $\phi_{1,2}$ are the $Z_2$-odd \textit{inert} doublets, $\langle \phi_{1,2} \rangle  =0$, and $\phi_3$ is the one $Z_2$-even \textit{active} doublet, $\langle \phi_3 \rangle =v/$\begin{scriptsize}$ \sqrt{2} $\end{scriptsize} $ \neq 0$, which plays the role of the SM Higgs doublet, with $h$ being the SM Higgs boson and $G^\pm,~ G^0$ the would-be Goldstone bosons. 
Note that the $Z_2$ charges assigned to each doublet are according to the $Z_2$ generator: odd-$Z_2$ charge to the inert doublets, $\phi_1$ and $\phi_2$, and even-$Z_2$ charge to the active doublet, $\phi_3$. Therefore, the symmetry of the potential is respected by the vacuum $(0,0,v/$\begin{scriptsize}$ \sqrt{2} $\end{scriptsize}$)$. 
The symmetry allows for introduction of CP-violation in the dark/inert sector leading to two inert charged states $S^\pm_{1,2}$ (a combination of $H^\pm_{1,2}$ states) and four neutral CP-mixed states, $S_{1,2,3,4}$ (and admixture of $H_{1,2},A_{1,2}$ states). The DM candidate, which is the lightest particle amongst the CP-mixed neutral fields from the inert doublets, is indeed stable due to the unbroken $Z_2$ symmetry. 

\section{Experimental signatures}

Scalars arising from non-minimal Higgs frameworks provide novel Higgs production/decay channels. 
At loop-level, extra charged scalars substantially contribute constructively or destructively to the $h \to \gamma \gamma/Z\gamma$ channels. 
Scalars from the inert doublets as viable DM candidates, can influence detectable properties of SM particles, such as the Higgs invisible decays, $h\to S_1S_1$, where $S_1$ is a scalar DM candidate with mass below $m_{h}/2$~\cite{Cordero-Cid:2016krd,Cordero:2017owj,Cordero-Cid:2018man,Cordero-Cid:2020yba,Keus:2019szx,Hernandez-Sanchez:2020aop}. 


\begin{figure}[hhh]
\begin{center}
\includegraphics[scale=0.9]{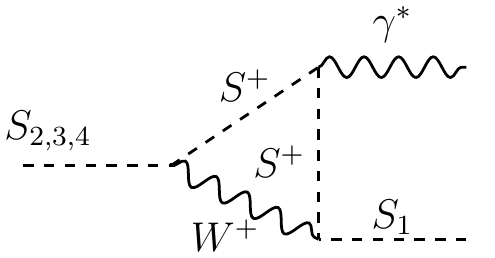}~~~
\includegraphics[scale=0.9]{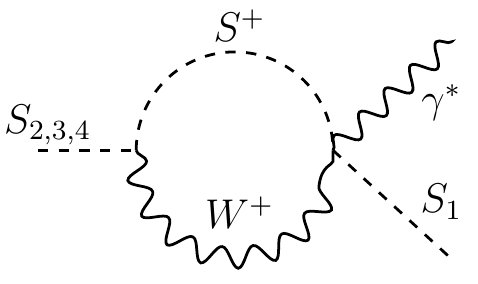}~~~
\includegraphics[scale=0.15]{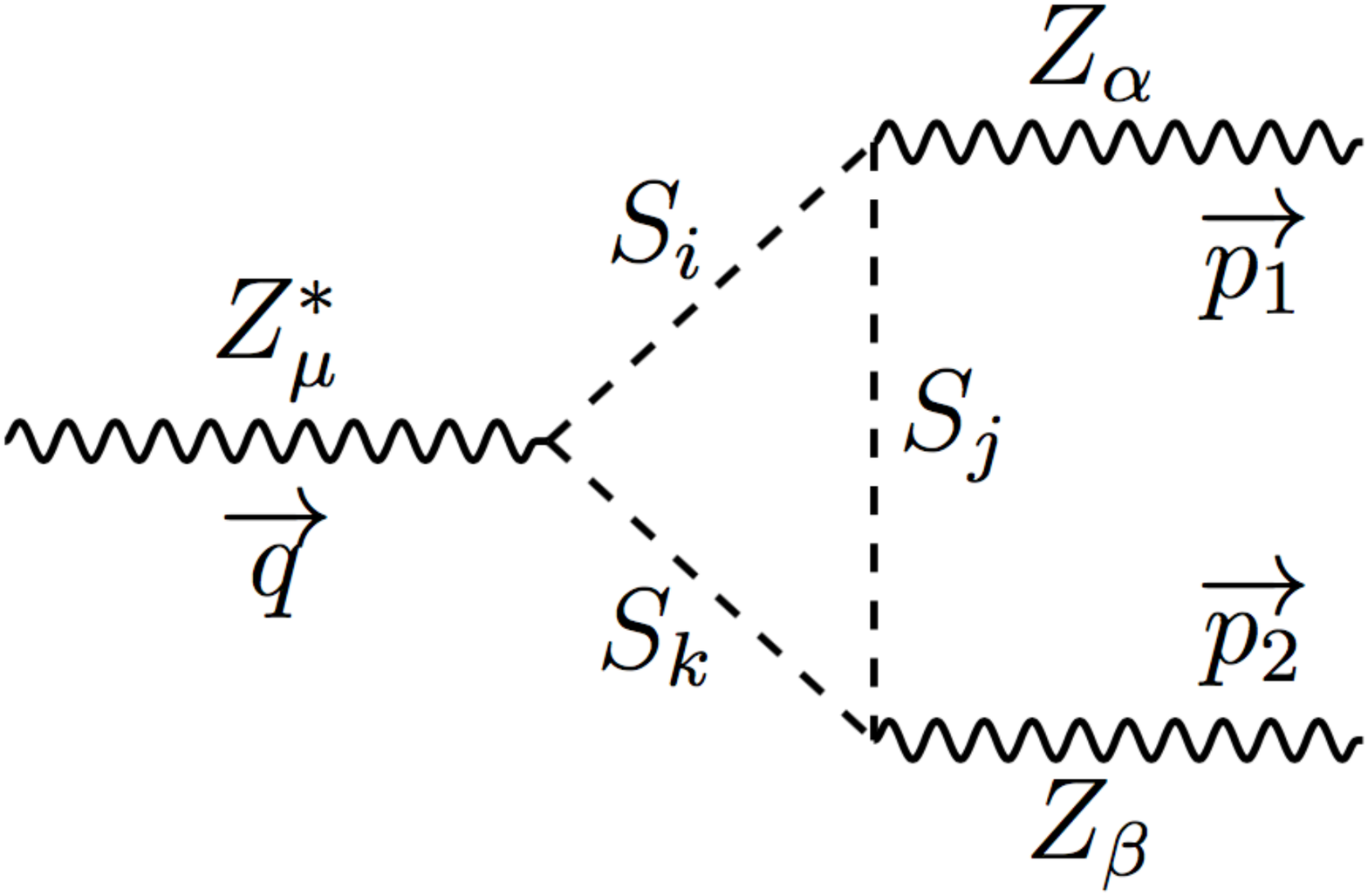}
\caption{Loop induced decays of the next-to-lightest scalar $S_{2,3,4}\to S_1\gamma^*$ (left and center) and contributions to the $ZZZ$ vertex with non-identical dark scalars $S_{i,j,k}$ in the loop (right).}
\label{fig:diags}
\end{center}
\end{figure}
Models with an extended inert sector have novel signatures such as loop induced decays of the 
next-to-lightest scalar, $S_{2,3,4} \to S_1 f \bar f$
mediated by both dark CP-odd and charged scalars, as shown in the left and center diagrams in Figure~\ref{fig:diags}.
This is a smoking gun signal of the 3HDM since it is not allowed in the
2HDM with one inert doublet and is expected to be important when $S_{2,3,4}$ and $S_1$ are close in mass.
In practice, this signature can be observed in the cascade decay of the 
SM-like Higgs boson, $h\to S_1 S_{2,3,4} \to S_1S_1 f \bar f$ into two DM particles and di-leptons/di-jets where $h$ is produced from gluon-gluon fusion or vector boson fusion.
Even though this signal competes with the 
tree-level channel $q\bar q\to S_1S_1Z^*\to S_1S_1f \bar f$, the resulting detector signature, $\cancel{E}_{T}\; f \bar f$, with invariant mass of $ f \bar f$ much smaller than $m_Z$, can potentially be extracted at the LHC already during Run 2 and 3. 
For example, the $S_{2,3,4}\to S_1\gamma^*$  and $\gamma^*\to e^+e^-$ case will give a spectacular QED mono-shower signal~\cite{Keus:2014isa}.

An extended inert sector accommodating dark CP-violation, with no contribution to the EDMs, can manifest itself in the active sector through contributions to the $ZZZ$ vertex, as shown in the right digram in Figure~\ref{fig:diags}, i.e. one-loop effects entering the cross section for $f \bar f \to Z^* \to ZZ$ at the LHC and future lepton colliders. 
Moreover, in an $f \bar f \to ZZ$ process, the polarisations of the $ZZ$ pair can be measured statistically from the angular distributions of their decay products. If the polarisations of the $Z$ bosons are known, one could define CP-violating observables (asymmetries) for the $ZZ$ state to test CP-violation at, potentially, the LHC by the end of its lifetime (after the HL-LHC runs) and at future lepton colliders such as the FCC-ee, ILC, CLiC or CEPC running at current design luminosities \cite{Cordero-Cid:2020yba}. 
%

\section*{Acknowledgements}
The author acknowledges financial support from Academy of Finland projects ``Particle cosmology and gravitational waves'' No. 320123
and ``Particle cosmology beyond the Standard Model'' No. 310130, and 
would like to thank the organisers of the Moriond QCD 2021 conference for creating a fantastic opportunity for lively scientific discussions.

\end{document}